 \documentclass[preprint2]{aastex}

\slugcomment{Submitted to \apj March 26, 2013}

\shorttitle{Solubility of iron in metallic hydrogen}
\shortauthors{Wahl et al.}

\begin{document}


\title{Solubility of iron in metallic hydrogen and stability
of dense cores in giant planets}


%\author[1]{Sean M. Wahl} 
%\author[2]{Hugh F. Wilson}
%\author[1]{Burkhard Militzer}
%\affil[1]{Department of Earth and Planetary Science, Department of Astronomy, University of California,
%    Berkeley, CA 94720, USA.}
%\affil[2]{Somewhere in Austrailia}

\author{Sean M. Wahl, Hugh F. Wilson\altaffilmark{1,2} and Burkhard
Militzer\altaffilmark{1}}
\email{swahl@berkeley.edu}
\affil{Department of Earth and Planetary Science, University of California,
    Berkeley, CA 94720, USA.}

    
\altaffiltext{1}{Department of Astronomy, University of California,Berkeley,
CA 94720, USA.}
\altaffiltext{2}{Virtual Nanoscience Laboratory, CSIRO Materials Science and
Engineering, Parkville, Victoria 3052, Australia.}

\begin{abstract}
The formation of the giant planets in our solar system, and likely a majority
of giant exoplanets, is most commonly explained by the accretion of nebular hydrogen and
helium onto a large core of terrestrial-like composition. The fate of this core has important
% reword terrestrial-like composition?
consequences for the evolution of the interior structure of the planet. It has
recently been shown that $\mathrm{H}_2\mathrm{O}$, MgO and $\mathrm{SiO}_2$ dissolve in liquid metallic 
hydrogen at high temperature and pressure. In this study, we perform {\it ab initio}
calculations to study the solubility of an innermost metallic core. We find
dissolution of iron to be strongly favored above 2000 K over the entire
pressure range (0.4-4 TPa) considered. We compare with and summarize the
results for solubilities on other probable core constituents. The calculations
imply that giant planet cores
are in thermodynamic disequilibrium with surrounding layers, promoting
erosion and redistribution of heavy elements. Differences in solubility
behavior between iron and rock may influence evolution of interiors, particularly
for Saturn-mass planets. Understanding the distribution of
iron and other heavy elements in gas giants may be relevant in understanding
mass-radius relationships, as well as deviations in transport properties from
pure hydrogen-helium mixtures.
\end{abstract}


%\keywords{ hydrogen-iron mixtures, {\it ab initio} simulations, giant planets, extrasolar planets}
%% and \citep commands to identify citations.  The citations are


%\section{Introduction}

Despite recent advances in computational methods improving understanding of
the hydrogen-helium dominated outer layers
\citep{mcmahon12,french12,militzer13,wilson10}, knowledge of the deep interior structure
of giant planets is limited.
Determining the size of a dense central cores in a giant planet is dependent upon
the model and equation of state used.
Current observational evidence yields recent estimates for present day core
mass of
$\sim$0$-$10 \citep{guillot05} and $\sim$14$-$18 \citep{militzer08} Earth masses for Jupiter, and
$\sim$9$-$22 \citep{guillot05} Earth masses for Saturn.
% use symbol for earth mass?
% might want to review whether this figure has been improved!
The Juno spacecraft, en route to Jupiter, will improve this constraint with more 
precise measurements of the giant's gravitational field \citep{helled11}. 
Meanwhile, the density profiles of Neptune and Uranus allow non-unique solutions 
for the compositional structure for much of the interior
\citep{guillot99b,guillot05}.

It has long been suggested
\citep{stevenson82a,stevenson82b}, that a portion of this dense material might be
redistributed in solution with hydrogen. As a result, erosion of a dense core
would cause it to shrink over the lifetime of the planet. Possible consequences of this process
are only beginning to be enumerated in evolutionary models
% reword enumerated?
\citep{chabrier07,leconte12,mirouh12}. The establishment of a gradient in
concentration of a heavy dissolved component may change the nature of
convection in a portion of the planets interior. This `double-diffusive'
is hypothesized to reduce the efficiency of heat transfer, thereby altering
the thermal evolution of the planet. Comprehensive understanding of the
process has been limited by the lack of knowledge of the solubility of various
phases in metallic hydrogen, as well as poor understanding of the scaling of
convective efficiency in the presence of competing gradients of composition
and temperature. In this study, we address the first issue for iron metal.

As a result of continuing discoveries by Kepler \citep{borucki10} and other exoplanet surveys, the
number of confirmed planets has climbed to over 800, the majority of which are
giants. This presents a growing sampling of planetary
mass-radii relationships that will be fundamental to understanding the evolution of
giant planet interiors. The range of mass-radius relationships observed for
exoplanets exhibit variation beyond those in the solar system. In some cases,
such as Corot-20b \citep{deleuil11}, the relationships may even defy explanation by
simple structural models. Redistribution of dense core material lowers the
heavy element content required to explain anomalously high observed densities.

\begin{figure}[h]
\epsscale{1.0}
\plotone{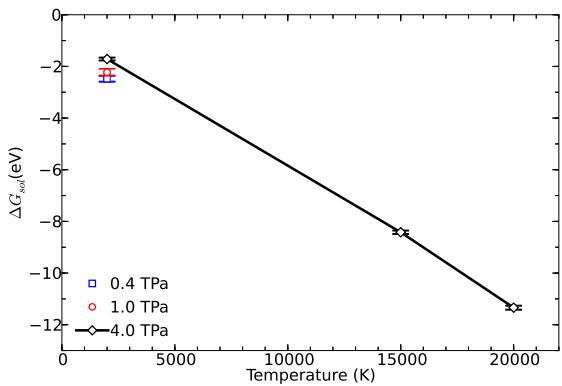}
\caption{Gibbs free enerby of solvation for solid Fe in liquid metallic
hydrogen. Negative values favor dissolution for a solute ratio of 1:256.}
\end{figure}

The favored model for gas giant formation
\citep{mizuno78,bodenheimer86,pollack96} relies on the early
formation of a large planetary embryo of critical mass to cause runaway
accretion of hydrogen and helium gas. A competing theory involves collapse of
a region of the disk under self-gravity, e.g. \citep{boss97}, but may
%this likely requires unrealistically high densities for most disks \citep{?} and 
have difficulty explaining significant enrichment of refractory elements \citep{hubbard02,guillot05}.
The immediate result of a core-accretion hypothesis is a planet with the
ice-rock-metal embryo residing at the center as a dense core, surrounded by an
extensive layer of metallic hydrogen and helium. The role of core erosion to
the subsequent evolution is a major source of uncertainty, but in principle, can explain shrinking of
cores to masses smaller than those necessary to form the planet under the
core-accretion hypothesis.

Core erosion in giant planets can be addressed by determining the solubility 
of analogous phases. Previous studies have considered an icy
layer of fluid and superionic $\mathrm{H}_2\mathrm{O}$
\citep{wilson12a,wilson13}, and a rocky layer consisting of
MgO \citep{wilson12b} and $\mathrm{SiO}_2$ \citep{gonzalez13}, which have been
shown to separate at relevant conditions \citep{umemoto06}. Assuming the
same gross distribution, elements as terrestrial bodies, the innermost core
component would be a dense, metallic alloy composed primarily of iron. 

{\it Ab initio} random structure searches
\citep{pickard09} demonstrate that iron remains in a hexagonal close packed (hcp) structure
remains stable up to pressures approaching Jupiter's center, $\sim$2.3 TPa, at
which point it undergoes a phase transition to face centered cubic (fcc)
structure. \citep{stixrude12} demonstrated a gradual decrease in this transition
pressure with temperature. Simulations of liquid hydrogen
\citep{militzer08,militzer13,mcmahon12} undergo a gradual transition from molecular to
metallic, which is complete by $\sim$ 0.4 TPa at low temperatures.

%Conditions on both sides of
%the hcp$\rightarrow$fcc transition are considered here, but not the higher pressure transition to
%body centered cubic. Consideration of liquid hydrogen is
%restricted to its metallic regime, above $\sim$3 Mbar and $\sim$1000 K.
%% Rewrite to not yet mention the current project?

\begin{table}
\begin{center}
\caption{Gibbs free energy of solvation for Fe in liquid H\label{solvation}}
\begin{tabular}{rrcc}
\tableline
\tableline
\phantom{(G}$P$\phantom{a)} & \phantom{(}$T$\phantom{))} &  Fe Phase & $\Delta G$ \\
(GPa) & (K)~ & - & (eV)  \\
\tableline
400 \phantom{0}  & 2000  & hcp  &  $-$2.2 $\pm$ 0.14    \\
1000\phantom{0}  & 2000  & hcp  &  $-$2.5 $\pm$ 0.12    \\
1000\phantom{0}  & 2000  & fcc  &  $-$2.3 $\pm$ 0.13    \\
%1000\phantom{0}  & 15000 & liq  &  $-$12.2 $\pm$ 0.20\phantom{0}   \\
4000\phantom{0}  & 2000  & fcc  &  $-$1.71 $\pm$ 0.056  \\
4000\phantom{0}  & 15000 & fcc  &  $-$8.42 $\pm$ 0.066  \\
4000\phantom{0} & 20000 & fcc  &  $-$11.34 $\pm$ 0.078\phantom{0} \\
\tableline
\end{tabular}
\end{center}
\end{table}

Compression of materials with modern experimental techniques can reach megabar
pressures, however the pressure-temperature conditions near Jupiter's core ($>$4 TPa) remain
inaccessible. Temperatures in shockwave experiments climb rapidly at high
pressures in comparison with planetary adiabats, while diamond anvil cell
experiments
are best suited for low temperatures. As a result, simulations based on ab 
initio theories are best suited for directly probing the conditions of gas 
giant interiors.

We performed density functional theory molecular dynamics (DFT-MD) simulations
to determine the energetics of a dissolution reaction, in which solid iron
dissolves in pure liquid hydrogen. We calculate a Gibbs free energy of
solvation:
\begin{eqnarray}
  \Delta G_{sol}\left(\mathrm{Fe}:256\mathrm{H}\right) &=&
  G\left(\mathrm{H}_{256}\mathrm{Fe}\right) \\ &-&
  \left[ G\left(\mathrm{H}_{256}\right) 
     + \frac{1}{32}G\left(\mathrm{Fe}_{32}\right) \right]\mathrm{,} \nonumber
\end{eqnarray}
where $G\left(\mathrm{H}_{256}\right)$ and $G\left(\mathrm{Fe}_{32}\right)$
are the Gibbs free
energies of a pure hydrogen liquid and solid or liquid iron.
$G\left(\mathrm{H}_{256}\mathrm{Fe}\right)$ is the Gibbs free energy of 1:256 liquid solution of iron
in hydrogen. We assume that analysis of a single low-concentration solution is
sufficient to determine the onset of core erosion, since the reservoir of
metallic hydrogen would be much larger than the core. This does not rule out
non-ideal effects of higher concentrations that might exist in a narrow,
poorly convecting layer at the top of a core. 
%Regardless, the increase in density 
%from inclusion of iron is expected to have a much stronger affect on the rate 
%of core erosion.

Free energy calculations require the determination of a contribution from
entropy, which is not determined from the standard DFT-MD formalism. To achieve
this, we adopt a two step thermodynamic integration method, as 
used in previous studies \citep{morales09,wilson10,wilson12a,wilson12b}. The method
requires integration of the change in Helmholtz free energy over an
unphysical, yet thermodynamically permissible, transformation between two
systems governed by potentials $U_a\left(\mathbf{r_i}\right)$ and 
$U_b\left(\mathbf{r_i}\right)$. We define a hybrid potential 
$U_{\lambda}=\left(1-\lambda\right)U_a+\lambda U_b$, where $\lambda$ is the
fraction of the potential $U_b\left(\mathbf{r_i}\right)$. The difference is
Helmholtz free energy is then given by
\begin{mathletters}
\begin{eqnarray}
  \Delta F_{b\to a} &\equiv& F_b - F_a \\ &=& \int_{0}^{1}{d\lambda\,\langle U_b\left(\mathbf{r_i}\right) -
  U_a\left(\mathbf{r_i}\right) \rangle_{\lambda}}
\end{eqnarray}
\end{mathletters}
where the averaging is over configurations generated
during simulations of the system governed by the hybrid potential. 
%A smooth function in
%$\lambda$ interpolating between these calculated values for 
%$\langle U_b - U_a\rangle_{\lambda}$ is integrated to
%yield $\Delta F_{a\to b}$.

To increase the efficiency, we calculated the Helmholtz free energy in two steps 
each involving an integral of the form of Eq. 2. We first find $\Delta
F_{\mathrm{DFT}\to \mathrm{cl}}$,
between the systems governed by DFT and classical pair potentials, which are
fit via a force-matching method \citep{izvekov04}. In the second step $\Delta
F_{\mathrm{cl}\to \mathrm{an}}$, between the pair potentials and a reference system with an
analytic solution $F_{\mathrm{an}}$. The Helmholtz energy for the DFT systems,
$F_{\mathrm{DFT}}=F_{\mathrm{\mathrm{an}}}+\Delta F_{\mathrm{DFT}\to \mathrm{cl}}+\Delta F_{\mathrm{cl} \to \mathrm{an}}$, may then be
compared directly. The first step requires DFT-MD simulations for a small
number of $\lambda$ values, while the second uses applies a much faster classical Monte Carlo
approach at numerous values of $\lambda$ to ensure a smooth integration.

\begin{figure}[h]
\epsscale{1.0}
\plotone{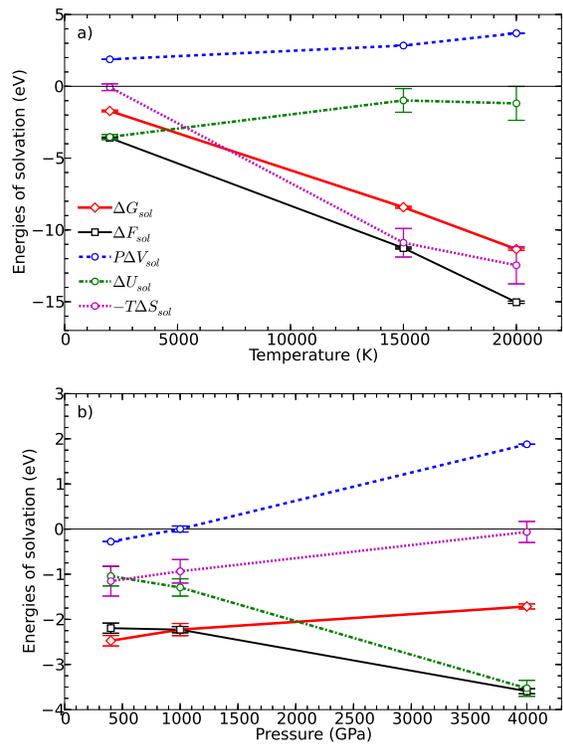}
\caption{Breakdown of $\Delta G_{sol}$ into contributions from: internal
energy,
$\Delta U_{sol}$, pressure effects, $P\Delta V_{sol}$, and entropic effects,
$-T\Delta S_{sol}$. Plots show variation with (a) temperature at P=4 TPa, and
(b) pressure at T=2000 K.}
\end{figure}

Finding a suitable reference system is essential to the method, as it allows
$\Delta F_{\mathrm{cl}\to \mathrm{DFT}}$ to be found with a small number of 
number of integration steps, and prevents solids from melting or transforming to a new
structure. For liquid systems, we integrate to a non-interacting
ideal gas, using classical two-body pair potentials as the intermediate
step. For solid iron, we integrate to an Einstein solid, a system of
non-interacting, 3-d harmonic oscillators. For the intermediate classical
solid system we use a 50-50
'mixture' of two-body and one-body harmonic oscillator potentials. Spring
constants for the Einstein terms are found from the mean-squared displacement 
of atoms from their ideal lattice sites during a DFT-MD simulation.

All simulations presented here were performed using the Vienna {\it ab initio}
simulation package (VASP) \citep{kresse96}. VASP uses the DFT formalism utilizing
pseudopotentials of the projector augmented wave type \citep{blochl94} and the exchange-correlation
functional of Perdew, Burke and Ernzerhof \citep{perdew96}. The iron
pseudopotential treats a $[\mathrm{Mg}]\mathrm{3pd}^6\mathrm{4s}^2$ electron
configuration as valence states, and a 2$\times$2$\times$2
grid of k-points is used for all simulations. Simulations on hydrogen 
and the solution were performed with a 900 eV cutoff energy for the plane wave expansion, while a 300
eV cutoff was used for iron. A time step of 0.2 fs was used for all liquid simulations, a
0.5$-$1.0 fs time step  was used for high and low temperature iron simulations 
respectively. The $\Delta G_{sol}$ results were confirmed to be well-converged
with respect to the energy cutoff and time step. 
Prohibitively long simulation times required that convergence with respect to
finer k-point meshes be verified over a subset of configurations generated by
a simulation with a 2$\times$2$\times$2 grid.
% this sort of gets repeated later on

% should add details about iron pseudopotential

%Solid iron has been shown to have an hcp structure up to $\sim$2.3 TPa, at which point
%it undergoes a phase transformation to fcc \citep{pickard09}. Phonon
%dispersion calculations demonstrate that the hcp and fcc structures remain
%stable at high temperatures \citep{stixrude12}.
%\vspace{12pt}

\begin{figure}[h]
\epsscale{1.0}
\plotone{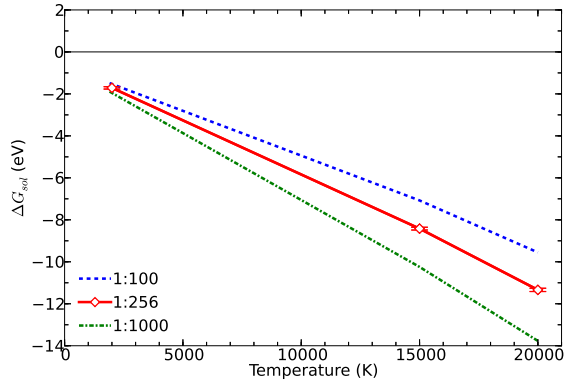}
\caption{Shift in $\Delta G_{sol}$ from a system with and Fe:H ratio of 1:256 to 1:100 and
1:1000 in the low-concentration limit.}
\end{figure}

%\vspace{12pt}

Iron simulations assume an hcp or fcc structure within their respective
stability regimes \citep{pickard09,stixrude12}. We confirmed Fe to be solid up
to 20000 K at 4 TPa, and to be a liquid at temperatures as low as 15000 K at
1 TPa. We also confirmed that the Gibbs free energy favors hcp stability over fcc
at 1 TPa, though the difference is negligible for our subsequent analysis of
dissolution. We found 32 atom supercells to be sufficient for Fe simulation.
Finite size effects required that we use large 256 atom supercells for
hydrogen, to which one Fe atom was added for the solution. Cubic
supercells are used for fcc and liquid runs. In order to maintain the same
number of atoms for the hcp an orthogonal supercell defined the combination of
hexagonal unit cell vectors $\mathbf{a}$,$\mathbf{a}+\mathbf{b}$,and
$\mathbf{c}$.

% add details about hcp cell

Cell volumes at each temperature were determined by fitting a pressure-volume polytrope equation of
state to short DFT-MD simulations. The resulting DFT pressures were all within
0.1\% of the target value. Gibbs free energies were computed for the three 
systems, $\mathrm{Fe}_{32}$, $\mathrm{H}_{256}$
and $\mathrm{H}_{256}\mathrm{Fe}_1$, using the thermodynamic integration method with simulation 
times of 1.0 ps for $\mathrm{H}_{256}$ and $\mathrm{H}_{256}\mathrm{Fe}_1$ 
and 2.5$-$5.0 ps for Fe. $\mathrm{H}_{256}$ and $\mathrm{H}_{256}\mathrm{Fe}_1$
runs with
$\lambda =1$ were extended to 4.0 ns for precise calculations of the internal energy,
which allows for determination of the entropic component of the Helmholtz free
energy. The calculated energies and entropy are presented in Tab. I,
along with the density.
The Gibbs free energy of solvation, calculated using Eq. 1, is presented in
Tab. II for each pressure-temperature condition. A negative $\Delta G_{sol}$
implies that the Gibbs free energy of the solution is lower than that of the separated
phases. Therefore, dissolution is favored at a solute concentration
higher than 1:256. 
% concentration versus ratio?

We find dissolution of iron to be strongly favorable at conditions corresponding to
the interiors of gas giants. Fig. 1 shows the variation of $\Delta G_{sol}$
with temperature and pressure. $\Delta G_{sol}$ exceeds $-$10 eV per iron atom for
plausible temperatures of Jupiter's core. Dissolution remains favorable even
at temperatures far below those predicted by model adiabats
\citep{militzer13,militzer13b}. The energetics
are only weakly dependent on pressure, and  $\Delta G_{sol}$ becomes increasingly
negative with decreasing pressure. The solubility increases with a nearly
linear trend in T, 
yielding slope of $\sim$0.53 meV/K. As a result, solubility is favored
through the entire range of conditions considered, and likely the entire range
for metallic hydrogen regions of giant planets.

Fig. 2 shows a breakdown of the data into contributions by various
thermodynamic parameters. Included in the figure are: $\Delta F_{sol}$, $\Delta U_{sol}$,
$P\Delta V_{sol}$ and $-T\Delta S_{sol}$, respectively, the Helmholtz free energy, internal
energy, volumetric work and entropic contributions contributions to $\Delta
G_{sol}$. Note that $\Delta F_{sol}$, $\Delta U_{sol}$ and $P\Delta V_{sol}$
are calculated independently, while $\Delta G_{sol}=\Delta F_{sol}+P\Delta
V_{sol}$ and $-T\Delta S_{sol}=\Delta F_{sol}-\Delta U_{sol}$ are derived. 
The trend of solubility with temperature is dominated by the entropic term.
The high solubility at low temperatures is
reflected in the negative values of $\Delta U_{sol}$, indicating that the
mixed system is energetically favorable independently of the entropy term.

Our calculations neglect any interactions between iron atoms in solution, and
thus represent solubility in the low-concentration limit.
$\Delta G_{sol}$ can be related to the volume change associated with
the insertion of an iron of atom into hydrogen, as other contributions are
constant with respect concentration. It can be shown that results for
simulations with a 1:n solute ratio can be generalized to a ratio of 1:m using
%\begin{equation}
%  \Delta G_{sol}(1:m) = \Delta G_{sol}(1:256) - k_BT\log\left[
%    \frac{\left(V(\mathrm{H}_n\mathrm{Fe}) +
%    \frac{m-256}{256}V(\mathrm{H}_n)\right)^{m+1}
%    V(\mathrm{H}_n)^{256}}
%    {\left(V(\mathrm{H}_n)\frac{m}{256}\right)^m
%    (V(\mathrm{H}_n\mathrm{Fe})^{256 + 1}} \right]
%\end{equation}
%  \begin{mathletters}
%  \begin{eqnarray}
%  \Delta G_c &\approx& F_0(H_mFe)-F_0(H_m)-F_0(Fe) \\ && -\left[
%    F_0(H_mFe)-F_0(H_m)-F_0(Fe)\right] \nonumber \\
%    -\frac{ \Delta G_c}{k_BT} &=& 
%  \end{eqnarray}
%  \begin{displaymath}
%    \log \left\{ 
%    \frac{\left[V(\mathrm{H}_n\mathrm{Fe}) +
%    \frac{m-n}{n}V(\mathrm{H}_n)\right]^{m+1}
%    \left[V(\mathrm{H}_n)\right]^{n}}
%     {\left[V(\mathrm{H}_n)\frac{m}{n}\right]^m
%    \left[V(\mathrm{H}_n\mathrm{Fe})\right]^{n + 1}}
%   \right\}, \nonumber
%\end{displaymath}
%\end{mathletters}
  \begin{mathletters}
  \begin{eqnarray}
  \Delta G_c &\approx& F_0(H_mFe)-F_0(H_m)-F_0(Fe) \\ && -\left[
    F_0(H_mFe)-F_0(H_m)-F_0(Fe)\right] \nonumber \\
    -\frac{ \Delta G_c}{k_BT} &\approx& 
    \log \left\{ 
    \frac{\left[V(\mathrm{H}_n\mathrm{Fe}) +
    \frac{m-n}{n}V(\mathrm{H}_n)\right]^{m+1}}
     {\left[V(\mathrm{H}_n)\frac{m}{n}\right]^m
    } \right\}  \nonumber \\
    && -\log \left\{ \frac {\left[V(\mathrm{H}_n\mathrm{Fe})\right]^{n + 1}}
    { \left[V(\mathrm{H}_n)\right]^{n}} \right\},
  \end{eqnarray}
\end{mathletters}
where $\Delta G_c = \Delta G_{sol}(1:m)- \Delta G_{sol}(1:n)$, and
$V(\mathrm{H}_n)$ and $V(\mathrm{H}_n\mathrm{Fe})$ are the volumes for the
simulations of hydrogen and the solution respectively. 
Fig. 3 shows the shift of
$\Delta G_{sol}$ at 4 TPa at Fe concentrations of 1:100 and 1:1000. $\Delta
G_{sol}$ is decreased for higher concentrations, but not to an extent where
dissolution would become disfavored. At 20000 K
this difference between 1:100 and 1:256 is  $<$2 eV per iron atom, and
at 2000 K is smaller than uncertainty in calculated values of $\Delta
G_{sol}$.

The nature of the Fe-H system poses additional numerical challenges compared to other solutes
considered previously \citep{wilson10,wilson12a,wilson12b,gonzalez13}.
These can largely be attributed to the comparatively large change in volume
and electron density associated with the insertion of an iron atom into
metallic hydrogen. We found it
more efficient to determine cell volumes by fitting an equation of state to a
collection of MD simulations at constant volume, rather than performing
extended constant pressure simulations. 
%This, consequently, eliminated the contribution of the
%error in the PV term, which was most significant in previous studies. 
Finite size effects were also found to be more significant for the Fe-H
system, due to iron's relatively large volume and number of valence electrons. Fig.
4 shows the convergence of a difference in internal energy between H and H-Fe
for MD simulations with 128, 256 and 512 hydrogen atoms. We find 256 hydrogen
atoms to be necessary in contrast to the previous studies that required only 128.
The convergence with k-point grid resolution is also slower
than in previous studies, and presents the greatest uncertainty in this study.
MD calculations with a 3$\times$3$\times$3 k-point grid are prohibitively expensive. An
estimate of this error for the results presented here is obtained by
evaluating the internal energy over configurations sampled from a MD
trajectory with a 2$\times$2$\times$2 k-point grid.\linebreak Fig. 5 shows this
estimated correction of $\Delta G_{sol}$ for the k-point grid used. 
The shifts are on the order $\sim$1 eV per iron atom, but are
consistently negative for both quantities, leading to dissolution being more favorable. 

%\vspace{24pt}

\begin{figure}[h]
\epsscale{1.0}
\plotone{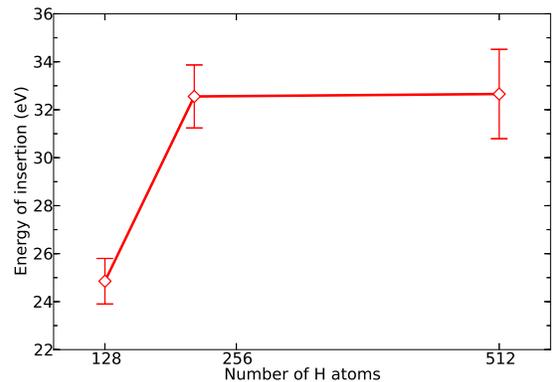}
\caption{Energy of insertion for a single Fe atom into supercells containing
128, 256, and 512 atoms. Finite size effects are significant
for $\mathrm{H}_{128}$, but are negligible within error for $\mathrm{H}_{256}$.}
\end{figure}

%\vspace{12pt}

With the results of previous studies
\citep{wilson12a,wilson12b,gonzalez13}, we can now present a comprehensive
picture for the solubility of all typical core materials in liquid
metallic hydrogen.
%% add once final figure is complete
%Fig. 7 shows calculated saturation limits for all
%considered phases,  $\mathrm{H}_2\mathrm{O}$, MgO, $\mathrm{SiO}_2$ and Fe, in
%comparison to thermal structures of Jupiter and Saturn. 
Dissolution is strongly favored for both iron and water
ice. However, for water the high solubility is attributed entirely to the
entropy, whereas iron has a favorable internal energy component that favors
dissolution at low temperatures.
Both phases are found to be soluble throughout the entire metallic
hydrogen region of both Jupiter and Saturn. The rocky components, MgO and
$\mathrm{SiO}_2$, have more moderate solubilities, with $\mathrm{SiO}_2$ being slightly higher. The saturation
curves are, however, less steep than the adiabats for Jupiter and Saturn. As a
result, solubility is favored for Jupiter's core, but the rocky components of
Saturn's core may be stable given the present uncertainty in the planet's
adiabat.

\begin{figure}[h]
\epsscale{1.0}
\plotone{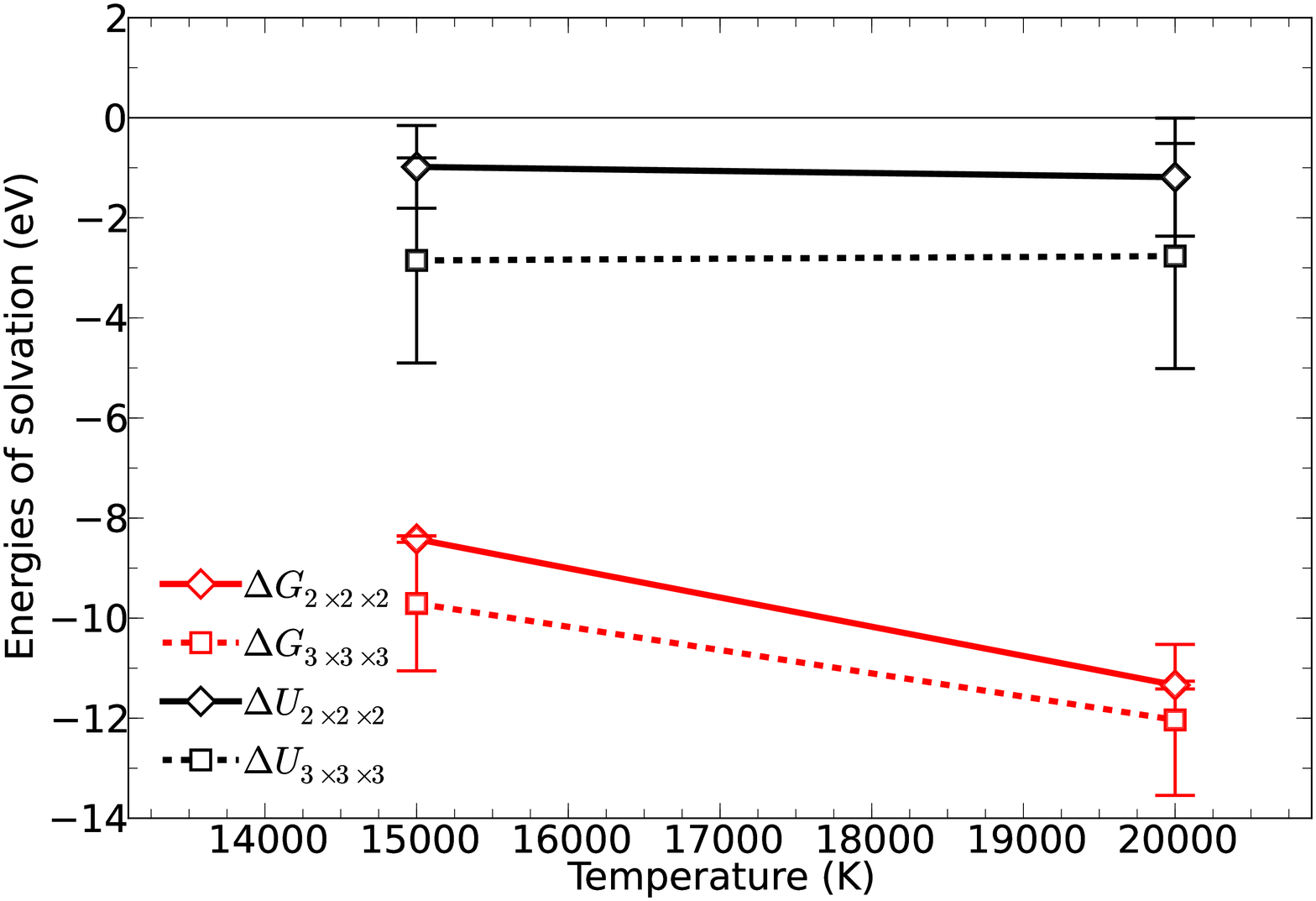}
\caption{Estimated corrections to $\Delta G_{sol}$ and $\Delta G_{sol}$
coarseness of k-point grid used in DFT-MD runs. DFT calculations with a $3\times
3\times 3$ k-point grid were performed sampling a trajectory generated by an MD
simulation with a $2\times2\times 2$ k-point mesh. }
\end{figure}

The rocky components are found to be
stable at lower pressures, approaching the metallic transition. If Mg and Si
were convected upwards with sufficient concentration, they may precipitate, while
Fe and $\mathrm{H}_2\mathrm{O}$ would remain in solution, at least to the molecular-metallic
transition. 
%The presence of a significant dissolved component at shallow depths
%might lead to a lowering in the molecular-metallic transition pressure. 
%Since this transition is linked to significant changes in material properties
%\citep{french12},
%its depth within gas giants may beimportant for their evolution. A lower
%transition pressure might also allow stabilization of metallic hydrogen within
%ice giants, where observations neither require or rule out its presence.
%A lower transition pressure would also aid experimental detection of hydrogen
%metallization.
The presence of a significant dissolved component at shallow depths may have
consequences for the density profile and transport properties of
hydrogen, which influence thermal structure and magnetic field
generation.

Core erosion is thermodynamically favorable in gas giant planets,
with the possible exception of smaller, cooler planets, like Saturn. For these planets, the
outer icy layers are soluble, but the rocky layers may not be. The innermost
iron component, though soluble, would be isolated from reaction with hydrogen.
This might allow Saturn to have a larger, less eroded core than
Jupiter, a result consistent with current observational constraints.
Nevertheless, our results imply
that confirmation of a massive core for Jupiter would support the
core-accretion model over gravitational collapse. While erosion of such a core
may be slow due to inefficient double-diffusive convection
\citep{stevenson82a,chabrier07,leconte12,mirouh12}, settling of dispersed
refractory material to form a core is inconsistent with our results. Late formation of a core
would require a large amount material from captured planetesimals
surviving descent to the planet's center.

It may be possible to attribute some
emerging trends in exoplanet mass-radius relationships to the difference in
solubilities between rock and ice, or rock and metal. However, as we have
shown, such thermodynamic differences are likely to only be significant in
smaller, cooler planets, where redistribution of dense material by
double-diffuse convection would be least efficient. The energetics of the
dissolution reaction should be insignificant compared to the role of density
in the redistribution of dense material. The work required to raise an iron of atom to
the molecular-metallic transition is on the order of 1000 eV, whereas the
contribution from the dissolution reaction is $\sim$1$-$10 eV. We conclude
that the process of core erosion is thermodynamically consistent with ab
initio simulations of the relevant materials, and its
significance warrants close consideration in future models of giant planet
evolution.

\acknowledgments

This work has been supported by NASA and NSF. Computational resources
at NAS and NCCS were used. We also acknowledge Dave Stevenson and Felipe
Gonzalez for helpful discussions.

\appendix

%% thebibliography produces citations in the text using \bibitem-\citep
%% to the KEY in the \citep commands (see the first section above).
%% place of the \citep commands.



\bibliographystyle{apj}

\begin{deluxetable}{rrlcrrrrr}
\tabletypesize{\scriptsize}
\tablecolumns{9}
\tablewidth{0pc}
\tablecaption{Thermodynamic parameters derived from DFT-MD simulations.\label{data}}
\tablehead{ \colhead{P} & \colhead{T} & \colhead{System} & \colhead{Phase} &
\colhead{$\rho$} & \colhead{F} & \colhead{U} & \colhead{G} & \colhead{S}\\
(GPa) & (K) & ~~~~~- &  - & ($\mathrm{g}/\mathrm{cm}^3$) & (eV)~~~~ & (eV)~~~~
& (eV)~~~~ & ($\mathrm{k}_b/\mathrm{K}$) }
\startdata
400   &  2000   &  $\mathrm{Fe}_{32}$     &  hcp  &  14.408  &  $-$190.8(4)\phantom{0}   &  $-$154.4(0)            &  323.3(4)\phantom{0}     &  211.(4)              \\
.     &  .      &  $\mathrm{H}_{256}$    &  liq  &  1.2709  &  $-$415.6(5)\phantom{0}   &  $-$228.(0)\phantom{0}  &  419.37(9)               &  1088.(2)             \\
.     &  .      &  $\mathrm{H}_{256}\mathrm{Fe}$  &  liq  &  1.5192  &  $-$423.8(1)\phantom{0}   &  $-$234.(0)\phantom{0}  &  427.1(8)\phantom{0}     &  1101.(6)             \\
1000  &  2000   &  $\mathrm{Fe}_{32}$     &  hcp  &  18.279  &  $-$12.4(0)\phantom{0}    &  18.1(2)                &  1000.8(5)\phantom{0}    &  177.(1)              \\
.     &  .      &  $\mathrm{H}_{256}$    &  liq  &  1.8916  &  23.0(3)\phantom{0}       &  175.9(6)               &  1425.6(6)\phantom{0}    &  887.(3)              \\
.     &  .      &  $\mathrm{H}_{256}\mathrm{Fe}$  &  liq  &  2.2534  &  20.4(1)\phantom{0}       &  175.(2)\phantom{0}     &  1454.7(1)\phantom{0}    &  898.(3)              \\
1000  &  2000   &  $\mathrm{Fe}_{32}$     &  fcc  &  18.269  &  $-$10.35(0)              &  19.8(0)                &  1003.4(6)\phantom{0}    &  174.(9)              \\
.     &  .      &  $\mathrm{H}_{256}$    &  liq  &  1.8916  &  23.0(3)\phantom{0}       &  175.9(6)               &  1425.6(6)\phantom{0}    &  887.(3)              \\
.     &  .      &  $\mathrm{H}_{256}\mathrm{Fe}$  &  liq  &  2.2534  &  20.4(1)\phantom{0}       &  175.2(3)               &  1454.7(1)\phantom{0}    &  898.(3)              \\
%1000  &  15000  &  $\mathrm{Fe}_{32}$     &  liq  &  16.970  &  $-$506.2(1)\phantom{0}   &  12(7).\phantom{0(0)}   &  585.(2)\phantom{00}     &  49(0).\phantom{(0)}  \\
%.     &  .      &  $\mathrm{H}_{256}$    &  liq  &  1.6315  &  $-$2064.1(1)\phantom{0}  &  595.(4)\phantom{0}     &  $-$437.9(2)\phantom{0}  &  2057.(5)             \\
%.     &  .      &  $\mathrm{H}_{256}\mathrm{Fe}$  &  liq  &  1.9468  &  $-$2091.9(7)\phantom{0}  &  598.(7)\phantom{0}     &  $-$431.7(8)\phantom{0}  &  2081.(6)             \\
4000  &  2000   &  $\mathrm{Fe}_{32}$     &  fcc  &  28.374  &  754.91(4)                &  777.6(2)\phantom{0}    &  3365.97(1)              &  131.(8)              \\
.     &  .      &  $\mathrm{H}_{256}$    &  liq  &  3.6375  &  1392.3(8)\phantom{0}     &  1487.6(4)              &  4310.0(0)\phantom{0}    &  552.(7)              \\
.     &  .      &  $\mathrm{H}_{256}\mathrm{Fe}$  &  liq  &  4.3078  &  1412.3(8)\phantom{0}     &  1508.(4)\phantom{0}    &  4413.4(7)\phantom{0}    &  55(7).\phantom{(0)}  \\
4000  &  15000  &  $\mathrm{Fe}_{32}$     &  fcc  &  27.826  &  382.1(9)\phantom{0}      &  87(5).\phantom{0(0)}   &  3045.7(0)\phantom{0}    &  38(1).\phantom{(0)}  \\
.     &  .      &  $\mathrm{H}_{256}$    &  liq  &  3.3618  &  $-$392.9(3)\phantom{0}   &  1917.(3)\phantom{0}    &  2763.9(8)\phantom{0}    &  1787.(3)             \\
.     &  .      &  $\mathrm{H}_{256}\mathrm{Fe}$  &  liq  &  3.9865  &  $-$392.2(5)\phantom{0}   &  194(3).\phantom{0(0)}  &  2850.7(4)\phantom{0}    &  1807.(2)             \\
4000  &  20000  &  $\mathrm{Fe}_{32}$     &  fcc  &  27.550  &  176.(9)\phantom{00}      &  92(2).\phantom{0(0)}   &  2866.(8)\phantom{00}    &  43(2).\phantom{(0)}  \\
.     &  .      &  $\mathrm{H}_{256}$    &  liq  &  3.2731  &  $-$1284.5(8)\phantom{0}  &  208(1).\phantom{0(0)}  &  1957.2(6)\phantom{0}    &  1952.(6)             \\
.     &  .      &  $\mathrm{H}_{256}\mathrm{Fe}$  &  liq  &  3.8824  &  $-$1294.0(9)\phantom{0}  &  210(4).\phantom{0(0)}  &  2035.5(1)\phantom{0}    &  1972.(0)             \\
\enddata
%\tablecomments{Table \ref{data} is published in its entirety in the 
%electronic edition of the {\it Astrophysical Journal}.  A portion is 
%shown here for guidance regarding its form and content.}
\end{deluxetable}

\end{document}